# Shared Mobility in Berlin: An Analysis of Ride-Pooling with Car Mobility Data

Alexander Schmaus[1, 2, 3], Felix Creutzig[1, 2], Nicolas Koch[1, 3], Nora Molkenthin[3]

[1] Mercator Research Institute on Global Commons and Climate Change, Berlin, Germany
[2] Technical University Berlin, Berlin, Germany
[3] Potsdam Institute for Climate Impact Research, Germany

**Abstract:** In face of the threat of a climate catastrophe and the resulting urgent need for decarbonization together with the widespread emergence of the sharing economy, shared pooled mobility has been suggested as an alternative to private vehicle use. However, until now all of its real-life implementations have served a niche market, adjacent to taxi services. To better understand this discrepancy, as well as the potential of pooled mobility, we have here simulated and analyzed pooled mobility on the street network of Berlin with car trip data as input for ride requests. We measure the rate of sharable trips, the relative travel time of passengers, the average occupancy of the vehicles, the relatively driven distance compared to driving with a private vehicle. We observe that for requests in the city center of Berlin it is possible to serve all mobility requests currently done by car, with around 4700 vehicles. The travel time is around 1.34 higher than with a private vehicle, the vehicle's occupancy increases to 2.6. The driven distance is reduced by 65%. In the whole area of Berlin we observe that a ride-pooling system with 10000 vehicles can serve 60% of the trips. The travel time is 1.4 times higher than with a private vehicle, the occupancy gets three and the driven distance is reduced by 40%.

# 1. Introduction

The implementation of sustainable traffic is one of the key challenges of decarbonisation. The transportation sector emits 15% of the global greenhouse gas, out of which private vehicles are the largest emitter [1]. In Germany, the transport sector is responsible for 18% of all emissions, with private vehicles alone being responsible for 11% of all emissions [2]. Thus, decarbonisation is not possible without lowering the emissions from private vehicles. While most of the focus here lies on electrification of private vehicles, a reduction of private motorized mobility offers several additional benefits, such as the reduction of pollution, noise, and traffic congestions [3], [4].
In this analysis we thus focus on pooling similar rides as a means for overall traffic reduction. Ride-pooling offers a flexible and convenient alternative to line-based public transport. Several studies show that ride-pooling or shared pooled mobility could make a large contribution to lowering energy demand and increasing traffic sustainability [5], [6].

Furthermore, shared pooled mobility would increase the accessibility of public transport itself [7] .

The implementation of ride-pooling services could be realized significantly faster than expanding public transportation systems, especially rail systems. From an infrastructural site it only requires streets and vehicles, which are both usable without any further development. From the software site apps and routing/pooling algorithms are required. Both are already developed by several ride-pooling operators, like MOIA in the city of Hamburg [8]. In contrast to other public transportation systems, which use the street networks, like bus systems, ride-pooling is capable of using the flexibility of vehicles. Thus, ride-pooling maintains one of the biggest advantages of private mobility. But, despite this, there are no signs of a wider usage yet. Instead, ride-pooling is currently mostly operating in the pooled taxi niche.

In Berlin specifically a ride-pooling service operated by the Berliner Verkehrsgesellschaft was active from 2018 to 2022. It was available in the eastern parts of the area inside of the Berlin Ringbahn and used 4.423 stations. During the four years of operation around 1,85 million passengers were transported, the proportion of shared trips was around 67% and the client satisfaction reached around 97% [9]. These numbers were negatively influenced by the Corona epidemic starting in the year 2020. After expansion of the exceptional permission (the so-called *Experimentierklausel*), the service was finished, although continuing operation of the service, even with an expansion to the whole area of Berlin, would have been possible. Instead, another on-demand ride-pooling service operating only in some of the eastern parts of Berlin [10].

Despite several promising tests and pilots, shared pooled mobility has not yet emerged as a widely utilized sustainable transport option. We can only speculate about the reasons behind this. Shared pooled mobility can only achieve acceptable delays in two scenarios. Either it operates close to a taxi service, in which small numbers of rides are occasionally pooled for a small reduction in fares. This is the niche, in which uber pool and MOIA typically operate, with fares slightly below taxi fares and travel times slightly above direct travel times, it tends to attract customers, who don't want to drive, but find public transport too cumbersome. The value of this scenario in the context of sustainability is questionable, with emissions gains from sharing quickly being eaten up by losses due to deadheading. The other scenario, in which acceptable delays are realistic, is at very high demands, so that sharing becomes naturally possible with small delays. To achieve such high demands it becomes necessary to be competitive in price and convenience with personal cars. This is the scenario, where shared pooled mobility has the potential to positively impact sustainability. However, when it comes to data, many studies have to resort to taxi data [11], [12] or data directly from ride-pooling providers [13], [14]. This, however, likely underestimates the total demand and distorts the spatial distribution.

Here, we thus use logged trips of private cars in Berlin as our data basis for an analysis of ride-pooling feasibility. We use the origin and destination points of this dataset as requests for a ride-pooling simulation. The service is simulated for a range of fleet sizes of the shared pooled mobility service with a focus on commuter trips made between 7 and 8 am. As networks we use a street network covering the whole area of Berlin and a smaller network covering only the city center of Berlin.

## 2. Methods

### 2.1. Ride-Pooling Simulation

The concept of ride-pooling is to bundle similar car trips into one vehicle of a ride-pooling fleet. By this, the occupancy of the vehicles is increased, while the driven distance and the number of necessary vehicles decreases. This concept is visualized in Fig 1.

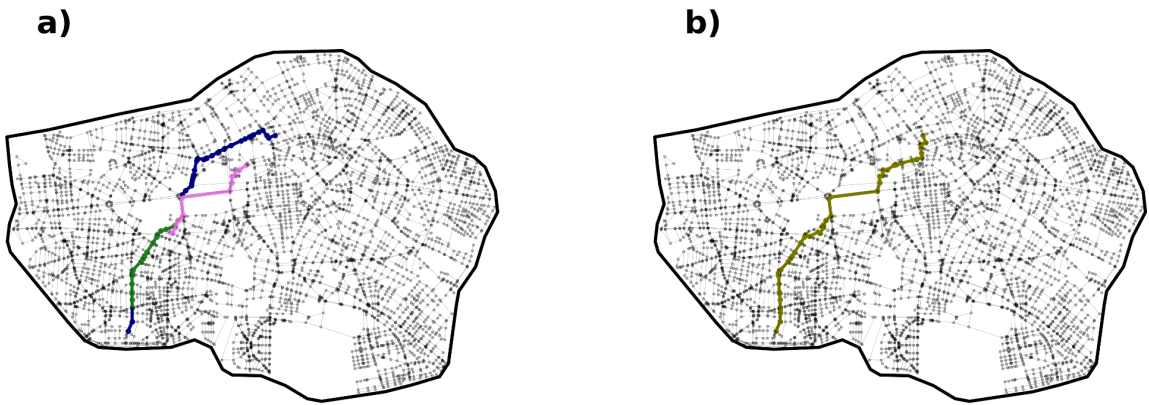

*Fig 1. Concept of ride-pooling. 1a) shows three individual car trips. To transport the five passengers, five vehicles are required. 1b) shows a possible pooling strategy. By this, only a single vehicle is required. The car trips are part of the INRIX Dataset and thus, real car trips from Berlin. The pooling strategy was determined by the ride-pooling simulation.*

To analyze ride-pooling systems we use an agent-based ride-pooling simulation [15]. The street networks of Berlin, on which this simulation is executed, are created with OpenStreetMap [16]. We use two networks: The first network includes the whole area of Berlin and has 10405 stops (c.f. Fig 2a). In the city center of Berlin, the public transport fare zone A, we use a denser network with 4696 stops (c.f. Fig 2b).

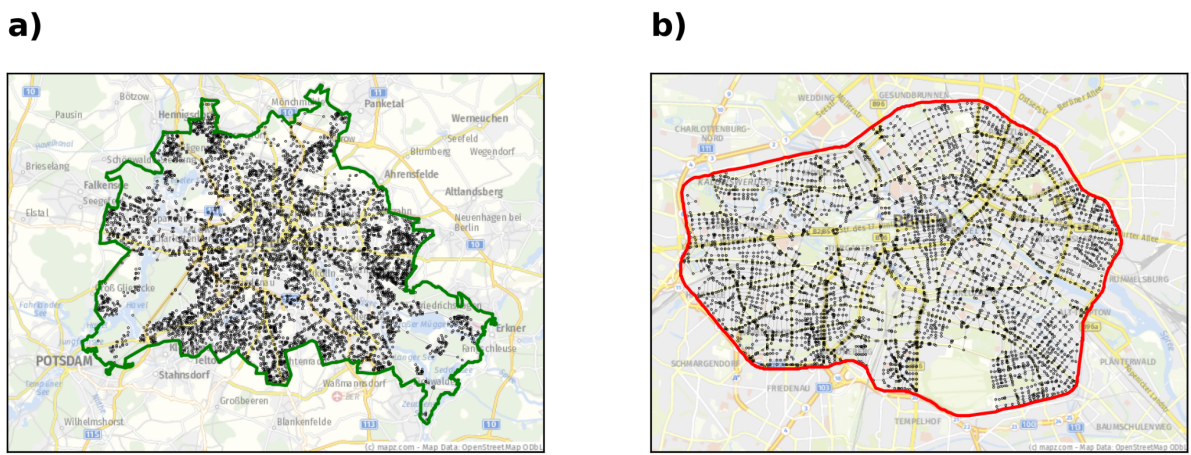

*Fig 2. Stop networks used in this work. 2a) shows the network covering the whole area in Berlin with 10,405 nodes. 2b) shows the network in the city center of Berlin (4696 nodes). The background image showing the map of Berlin was downloaded from [17].*

Before the simulation starts, an initial position is determined for every vehicle of the ride-pooling fleet, by uniformly drawing positions from the set of all stops. The effect of this strategy is discussed in chapter 4.3. Apart from the network and the initial positions, the following simulation parameter are important:
- Fleet size or number of vehicles
- Maximum pick-up or waiting time
- Maximum delivery delay
- Average speed

The fleet size defines the number of vehicles the ride-pooling service. The number of seats (capacity) can be defined separately for each vehicle or a general number is used. The maximum pick-up time is defined as the longest permitted waiting time for a passenger. If the maximum pick-up time is exceeded by every vehicle, the request is rejected. Similarly, the maximum delivery delay is defined as the longest permitted excess of the pooled trip duration over the direct driving time. If it is exceeded for a ride, the request will also be rejected. Furthermore, an average speed of the vehicles is defined. The chosen parameters heavily influence the functionality and efficiency of a ride-pooling service.

During the execution, the simulation processes the requests one by one. For each request, the simulation determines for each vehicle how much additional distance the vehicle must drive to process the request, while maintaining the time restrictions. The request is then assigned to that vehicle, which could serve the requests with respectively minimal additional distance.

To run the ride-pooling simulation, service requests are required. In this paper we generate the service requests from real car-trip demand in Berlin. This is further explained in section 2.3, after the introduction of the dataset in section 2.2.

## 2.2. Dataset

In Berlin, around 11.9 million trips were made inside of the city (internal traffic of the city of Berlin) each day in the year 2018 [18]. Of these 11.9 million trips, 18% were made by a car, resulting in 2.14 million vehicle trips within Berlin every day [19]. The temporal distribution of car trips is subject to strong temporal fluctuations. These are shown in Fig 3.

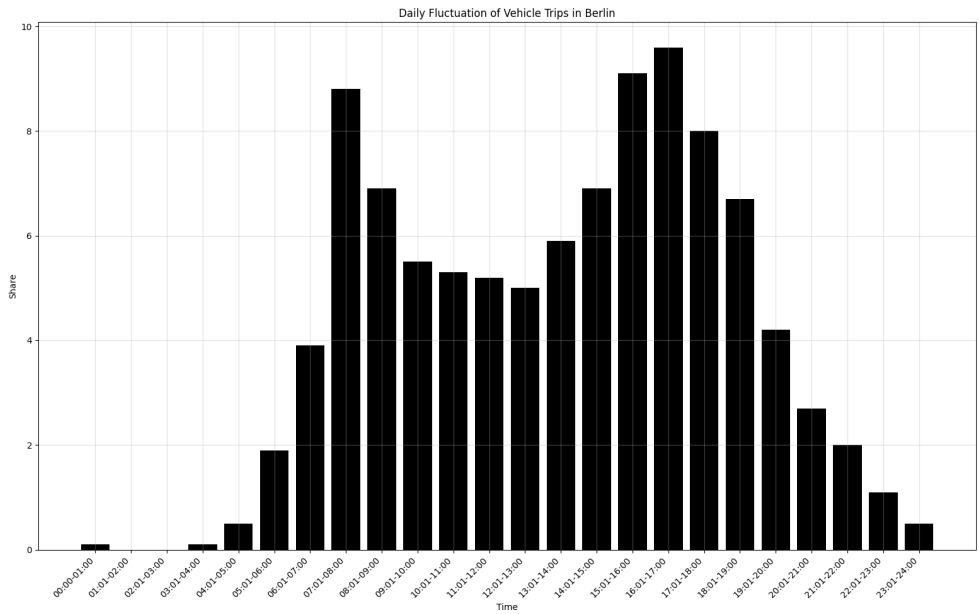

*Fig 3. Daily Fluctuation of car traffic in Berlin [19]. From 7 to 8am 8.8% of all car trips are made. This high number is formed by commuter traffic at this time.*

Since we primarily study commuter trips in this paper, only trips between 7 and 8am are considered in the following. This corresponds to the time with the highest traffic volume in the morning in Berlin. 8.8% of all car trips are made in this time slot. Thus, 188,320 car trips are made between 7 and 8 am in Berlin.

The dataset we use in this work is made available by the commercial data provider INRIX. Originally it contains 34,208,544 unique data points, including car trips starting from Berlin, ending in Berlin or crossing Berlin. The data was collected in 2017. It contains GPS data from private and commercial vehicles. All trips with origin and/or destination outside of Berlin are excluded for this work.

To use the trips as requests in the ride-pooling simulation, the origin and destinations are mapped to the stops in one of the two ride-pooling stop networks used in this work (c.f. Fig 2a. and Fig. 2b). For the origin and destination of every trip, the nearest stop in the respective network is searched. After that, all requests with the same origin and destination are removed. By this we end up with 769,650 trips inside of Berlin and 158,329 requests in the center of Berlin. By mapping the original origin and destination points to stops a walking time for each passenger is introduced. For the whole area of Berlin we get an average walking time of 255m, in the city center every customer has a walking time of around 137m.

In order to get near to the 188000 trips made between 7 and 8 am in Berlin, we divide the trips quarterly. Thus, we get four trip sets for the whole area and the city center of Berlin:

| Months | Abbreviation | Number of Trips Berlin | Number of Trips Berlin City Center |
|---|---|---|---|
| January, February, March | Q1 | 187329 | 37462 |
| April, May, June | Q2 | 193870 | 39570 |

| July, August, September | Q3 | 192170 | 40830 |
| October, November, December | Q4 | 196281 | 40467 |

*Tab. 1. Request sets formed from the INRIX dataset in order to represent daily traffic in Berlin and in the city center of Berlin between 7 and 8am. Only trips with origin and destination in Berlin or in the city center of Berlin are considered.*

From the requests set containing all requests within Berlin, we determine that the overall average speed of the vehicles is 25.32 km/h. The average travel time for all requests is 15 minutes and 23 seconds. In the requests set containing the trips inside of the center of Berlin, we observe an average speed of 18.26 km/h for all requests and an average travel time of 14 minutes. These measurements are later used to define some of the parameters used in the ride-pooling simulation. An overview of the most important parameters for Berlin and the center of Berlin is given in Tab. 2.

|  | **Berlin** | **Berlin City Center** |
|---|---|---|
| **Average Speed [km/h]** | 25.3 km/h | 18.3 |
| **Duration [min]** | 15.38 | 14 |

*Tab. 2. Average values of the trips used to define some of the parameters of the ride-pooling simulation.*

## 2.3. Definition Simulation Parameters

With the dataset it is now possible to define the simulation parameters described in section 2.1.
Studies show that passengers are willing to accept 50% longer trip durations with public transport compared to using a private vehicle [20]. Thus, we set the maximum delivery delay to be the half of the average trip duration measured from the data. Busses, subways and metro Lines in Berlin mostly have a 10 minute interval. We copy this as a maximum waiting time for the ride pooling service. The speed of ride pooling vehicles is constrained by the speed of general traffic, which between 7 and 8 am is around 18 km/h in the city center and 25 km/h in Berlin (see Tab. 2). Further we assumed that walking to the station and walking from the station to the desired location takes on average one minute each in the city center of Berlin and two minutes in the whole area of Berlin. In Tab. 3 an overview of the selected and fixed parameters is given:

|  | **Berlin** | **Berlin City Center** |
|---|---|---|
| **Number of stops** | 10405 | 4696 |
| **Average vehicle speed [km/h]** | 25.3 | 18.3 |
| **Maximum waiting time [min]** | 6:00 | 6:00 |

| Maximum delivery delay [min] | 7:40 | 7:00 |
|---|---|---|

*Tab. 3. Overview of the fixed parameters for the ride-pooling simulations in Berlin and the city center of Berlin.*

## 2.4. Pooling Characteristics

To evaluate the efficiency and functionality of a simulated ride-pooling system and to make the results comparable to other research, we decided to measure the following characteristics:

- **Share of serviced requests**: Depending on system parameters and fleet configuration, typically a fraction of requests can not be served within the service quality constraints. The share of serviced requests measures how many of the original requests were successfully delivered at their destination within the selected constraints. The rejected requests are assumed to continue to use private vehicles in some calculations.
- **Relative travel time**: The relative travel time measures how long it takes passengers to get to the desired location using the ride-sharing service compared to traveling with their own car. It includes the driving time and the waiting time at the station and walking times from and to a station. We use the travel times specified in the dataset as the time measurement for the use of the private vehicle. For rejected requests the original travel time from the dataset is used.
- **Relative driven distance**: The relative driven distance measures the proportion of the actual driven distance when using private vehicles (from the dataset) and the distance driven by the ride-pooling vehicles in the simulation. If the value is smaller than one, less distance is driven with the ride-pooling service. Equal to the relative travel time, the original distance from the dataset is used for rejected requests.
- **Empty mileage share**: Proportion of the driven distance where the vehicle is empty compared to the complete distance driven by the ride-pooling vehicles. This measurement is independent of the driven distance of the car trips.
- **Average vehicle occupancy**: The average occupancy measures the average number of customers simultaneously in each vehicle.
- **Number of empty vehicles**: The number of vehicles of the fleet, which were not used during operation of the system.

The selection of values was influenced by [21].

# 3. Results

First, we consider the results of the simulations within the city center of Berlin. We ran simulations with different fleet sizes for every quarter. The measurements of the characteristics for the simulation results with the request set Q1 are shown in Fig. 4.

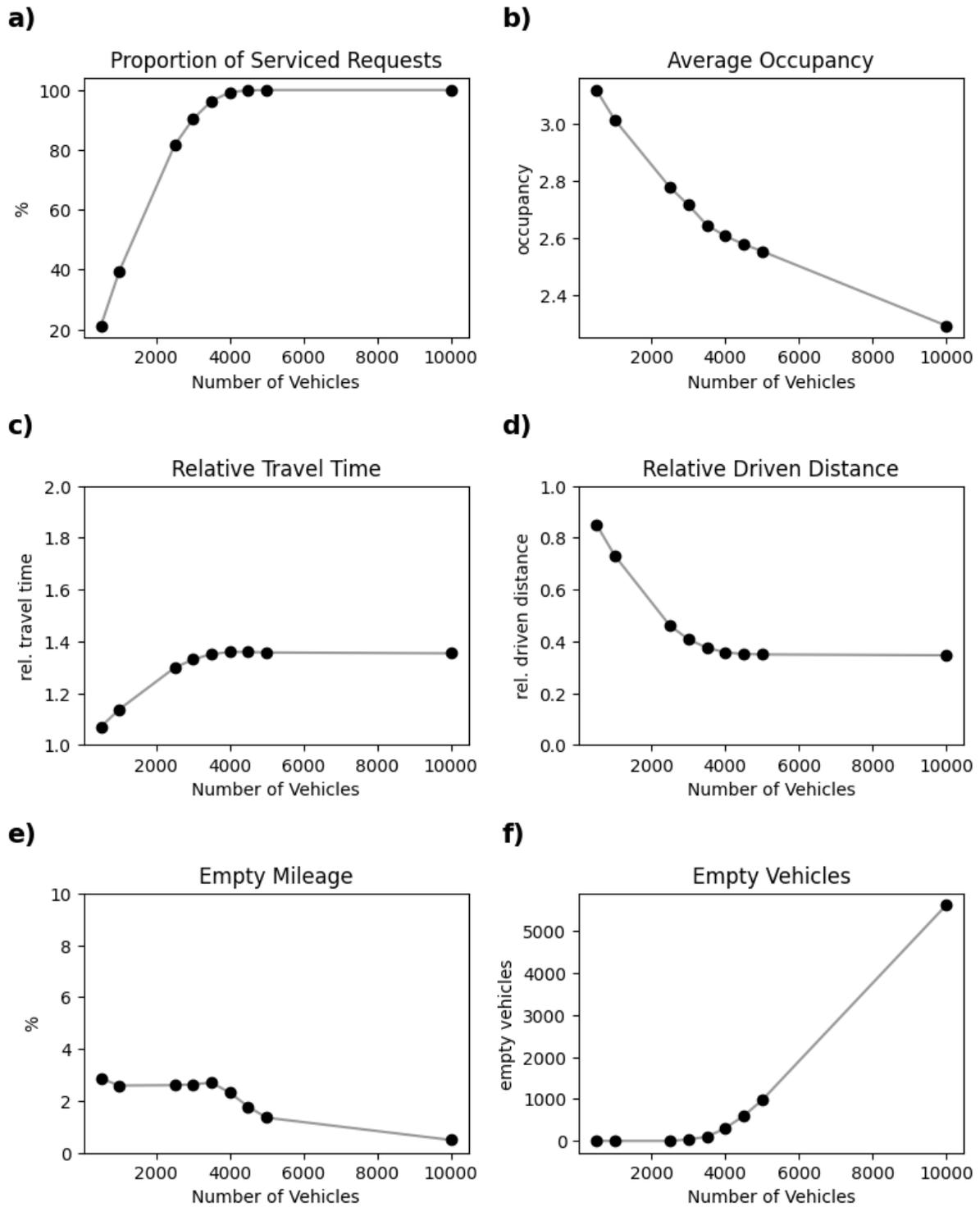

Fig. 4. Results for Q1 with differently sized ride-pooling fleets. 4a) shows the proportion of serviced requests for different fleet sizes. We observe that the lowest number of vehicles, which is capable of serving all requests, is around 4500 vehicles. 4b) shows the progress of the average occupancy. 4c) shows the progress of relative travel time and 4d) the progress of the relative driven distance. 4e) shows the share of empty mileage. 4f) shows the number of unused vehicles.

We observe a change in the behavior of all six characteristics around a fleet size of 4500. At this point the fraction of serviced request approaches 100 % (Fig.4a), the slopes of the decline in occupancy and the empty mileage change abruptly (Fig.4b,e), relative travel time and distance reach constant values (Fig.4c,d) and the number of idle vehicles increases (Fig.4f).

We interpret the regime change as a saturation occurring when the *minimal required fleet size* is reached. We determined this fleet size for every quarter, the results are shown in Tab. 4. At the minimal required fleet size, the decline in occupancy sharply changes as more and more vehicles remain empty (Fig. 4b). Fig. 4c) shows that the relative travel time increases for an increasing number of vehicles. This is because rejected trips are counted as using private vehicles and thus have a relative travel time of 1. As the fraction of pooled trips increases, so does the relative travel time, until it reaches the maximum permitted pooled travel time of 1.34. After the lowest number of vehicles capable of serving all requests is reached, the relative travel time remains stable. This is due to the implementation of the dispatcher algorithm, which prefers used vehicles over empty vehicles. The same effect holds for the relative driven distance. It decreases until finding the lowest number of vehicles, visible in Fig 4d). 4e) shows the progress of the empty mileage. We see that empty mileage does not play an important role. 4f) shows how many vehicles of the service were unused. The occurrence of this phenomenon that even if not all requests are serviced some vehicles stay empty, is explained in chapter 4.3.

For the lowest number of vehicles capable of serving all requests, the characteristics of all quarters are shown in Tab. 4.

| Quarter | Number of Vehicles | Relative Travel Time | Relative Driven Distance | Empty Mileage Share | Average Vehicle Occupancy | Empty Vehicles |
|---|---|---|---|---|---|---|
| Q1 | 4500 | 1.36 | 0.35 | 1.7% | 2.58 | 597 |
| Q2 | 4750 | 1.33 | 0.35 | 1.8% | 2.6 | 709 |
| Q3 | 5000 | 1.32 | 0.35 | 1.6% | 2.62 | 795 |
| Q4 | 4500 | 1.35 | 0.35 | 1.9% | 2.61 | 412 |
| **Average** | **4688** | **1.34** | **0.35** | **1.75%** | **2.6** | **628** |

*Tab. 4: Average characteristics observed with ride-pooling simulation experiments and the lowest number of vehicles which is capable of serving all requests.*

Due to the very high simulation times on the whole area of Berlin, the results are very limited compared to the results from the city center of Berlin. For the request set Q1 and the whole area of Berlin we get the results visualized in Fig. 5.

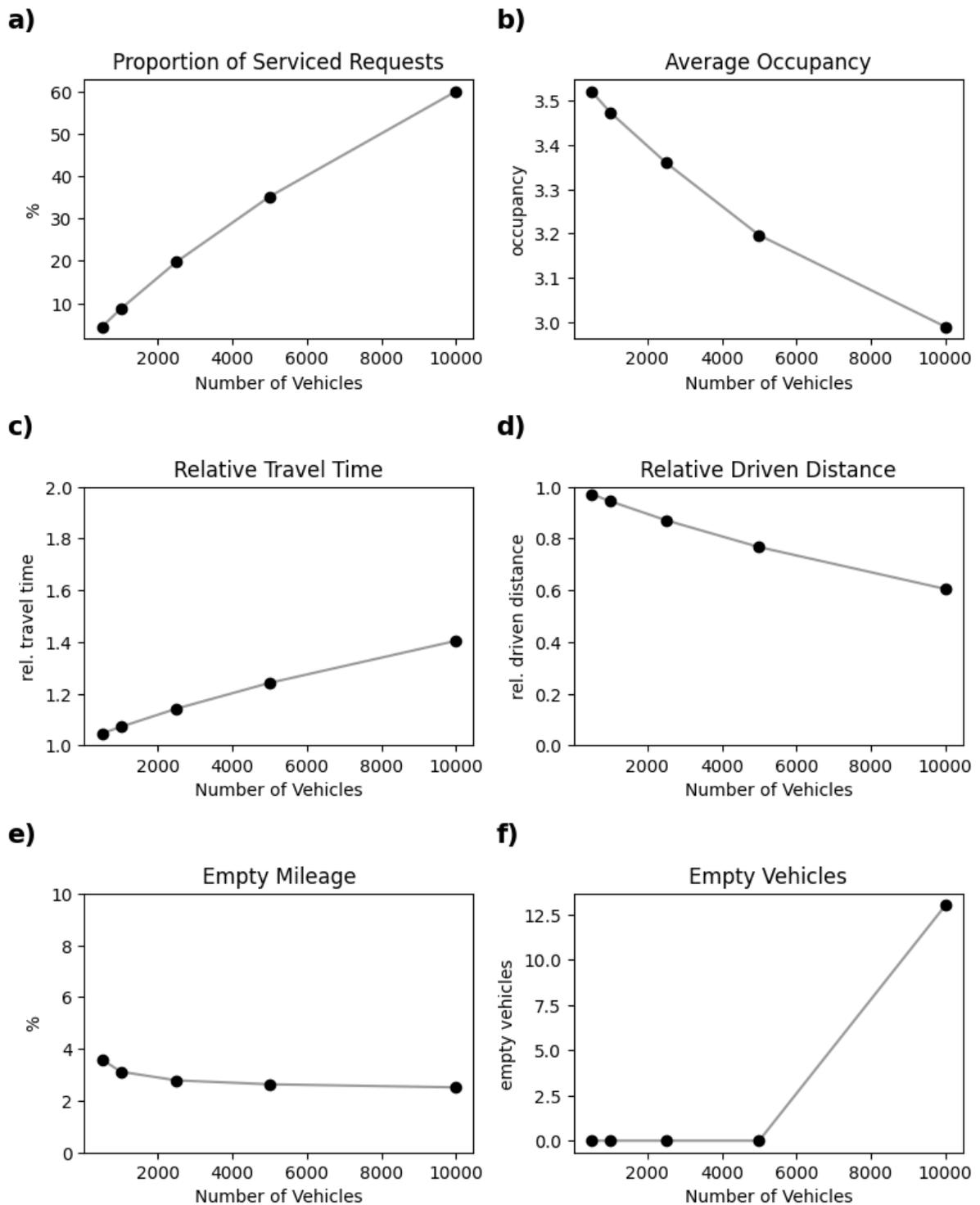

*Fig. 5. Progress of the characteristics in the whole area of Berlin for different fleet sizes. Due to the high simulation times we were not able to find a fleet size, which is capable of serving all requests.*

Fig. 5 shows that the characteristics of a ride-pooling service in Berlin are similar to the characteristics in the city center of Berlin. Fig. 5a) prompts that also for the whole area of Berlin a lowest number of vehicles exists, which is capable of serving all requests. A further discussion of these results is given in chapter 4.2.

# 4. Discussion

## 4.1. Characteristic Values Berlin City Center

Tab. 4. shows that the lowest number of vehicles capable of serving all car tips in the city center of Berlin is 4688.

Tab. 4 further shows that passengers take on average around 1.34 times as long to reach their destination with ride-pooling compared to private vehicle driving. For comparison, customers are 1.9 times slower compared to driving, when using public transportation[22].

The total driven distance is reduced by around 65%. This results in 65% less road traffic as well as $CO_2$-emissions. The entire motorized mobility demand is thereby met with 4500 vehicles instead of using 40000 private vehicles, a reduction by almost 90 %. The number gets even higher if it is considered that the ride-pooling vehicles can be used over the whole day.

The vehicle occupancy is around 2.6 and, thus, significantly higher than the current occupancy of 1.6 [19]. This is despite assuming each request to only account for one customer. Thus, in practice some trips in the INRIX dataset will have more than one customer. Between the average vehicle occupancy and the relative driven distance exists the following correlation (if all requests are accepted or the rejected request are ignored in the calculation of the relative driven distance):

$$average\ vehicle\ occupancy\ =\ \frac{1}{relative\ driven\ distance}$$

With the average vehicle occupancy equal to 2.6 and relative driven distance of 0.35 this formula does not hold. The reason for this is that in the ride-pooling simulation only shortest paths between two stops are chosen as driven paths. In the dataset this is not necessarily the case. Here, drivers for example choose another route to avoid traffic congestions or construction sites. Further it could be assumed that the data is noisy, meaning for example that drivers choose not the shortest path because they want to get to some other intermediate targets, for example to drop their children at schools.

If we calculate the relative driven distance not with the distance from the dataset but instead use the shortest paths between all origin-destination pairs, we get a new relative driven distance equal to 0.38, meaning that now less distance is relatively saved. If we insert the two values in the correlation formula we clearly see that it now roughly holds. That the values still differ is due to rounding effects in the calculation.

Empty mileage doesn't play an important role in the scenario and is only around 1.75%. This is due to the fact that the number of vehicles is as low as possible. At the same time, important aspects such as travel to the depot are ignored. The number of unused vehicles is reflected in more details in the next chapter.

## 4.2. Characteristic Values Berlin

In Fig. 5 the results for the whole area of Berlin are shown. Due to the high simulation and data processings times, we were only capable of simulating fleet sizes up to 10000 vehicles. 10000 vehicles are not enough to accept all requests but are already capable of serving around 60%. The average occupancy increases to three, the driven distance is reduced by around 40%. The relative travel time for 1000 vehicles is around 1.4, and thus higher than in the city center of Berlin. Like in the city center of Berlin, empty mileage can be ignored.

## 4.3. Optimal Fleet Size

As mentioned in chapter 3 we determined the lowest number of vehicles capable of serving every request for every quarter in the city center of Berlin. However, Fig. 2f. and Tab. 4 show that at this number a high amount of vehicles is unused. Even if not all requests are serviced, like with 3000 vehicles, a few vehicles are unused (c.f. Fig. 2a) and Fig. 2f)).

The phenomenon can be explained by the fact that the starting points of the rejected requests and the position of the unused vehicles are so different that the time constraints cannot be met. This is visualized in Fig. 6 .

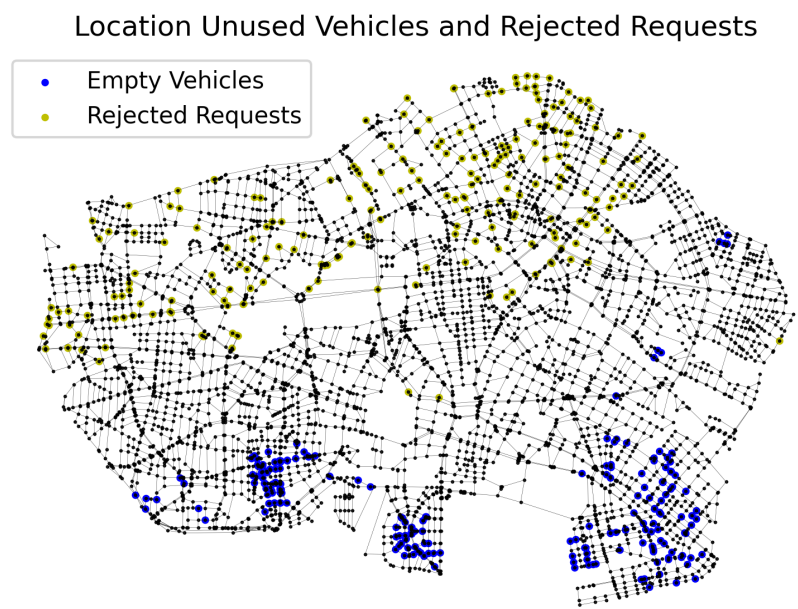

*Fig. 6. Location of empty vehicles and trip origins of rejected trips of a simulation with 4000 vehicles in the city center of Berlin. The empty vehicle locations are marked blue, while the trips origins of the rejected requests are marked yellow.*

Fig. 6 shows that the location of the empty vehicles are in the south of the city center of Berlin, while the origins of rejected trips are located in the northern parts of the network. Thus, if vehicles would drive from the south to the north to fetch customers, this would result in a waiting time restriction violation.

We furthermore see a correlation between the location of empty vehicles and nodes without any trip origin, describing why the empty vehicles stayed unused in the beginning of the simulation. This effect is visualized in Fig. 7.

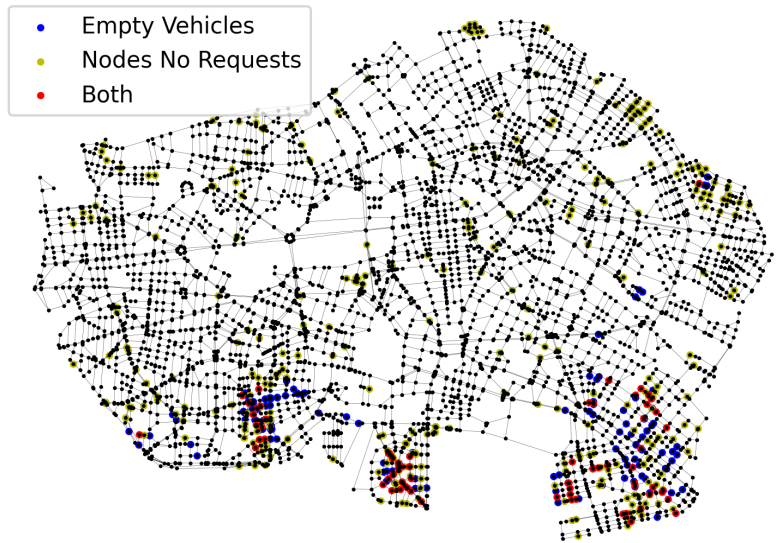

*Fig. 7: Locations of the unused vehicles (blue), nodes without any request origin (yellow) and nodes with unused vehicles and no request origin (red).*

This leads to the assumption that the initial locations could be chosen more effectively than by drawing positions uniformly from all nodes. This would require a sophisticated rebalancing algorithm to choose proper initial conditions of the vehicles [23].

## 4.4. Distortion of Results due to Decay Phase of Simulation and Vehicle Speed

Rejected requests occur because time restrictions cannot be met, or because no vehicle has free seats. Nevertheless, the average occupancy of the vehicles is never six (number of seats per vehicle). On the one hand, this is due to the fact that the cars start empty and are filled up over time. On the other hand it is due to the decay phase at the end of the simulation. Since only trips that start and end between 7 and 8am are taken into account, the number of requests at the end is smaller than at the beginning. Thus, the cars empty slowly reaching the end of the simulation. The progress of the occupancy of a single vehicle from a simulation in the city center of Berlin, with a fleet size of 500 vehicles, is shown in Fig. 7.

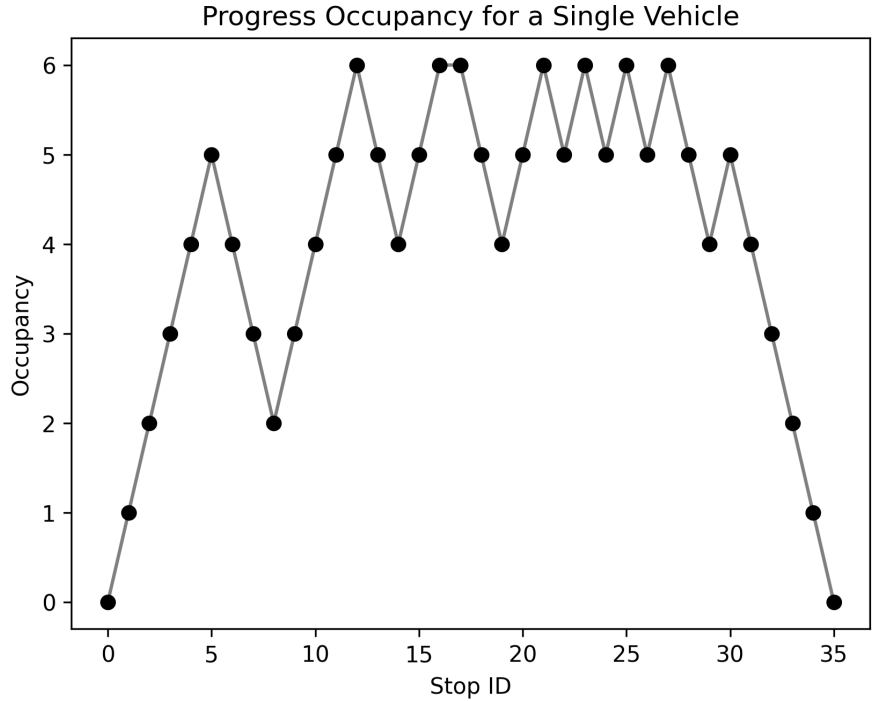

*Fig. 7. Development of the occupancy for a single vehicle in a simulation with a fleet size of 500 cars.*

The start-up time and decay time is independent of the simulation duration. This means that in case of a longer simulation time, the average occupancy of the vehicles increases. If the test used to create Fig. 7 is repeated with two hours instead of one, the average occupancy of the vehicles increases from 3.12 to 3.6. All results shown in Fig. 4, Fig. 5 and Tab. 4 are influenced by this behavior.

Secondly the relative travel time is strongly influenced by the vehicle speed used in the simulation, it can easily be increased or decreased by changing the average speed. We used the average speed we determine from the dataset as the average speed of the simulation. But, this is not necessarily the possible average speed of a potential ride-pooling service covering the two areas in Berlin we used as networks.

## 4.5. Cost Assessment and Comparison of Costs

In this section, we compare the total costs for car trips and the total costs for a ride-pooling system that replaces the individual trips. We use in this section an average value of 40000 travelers in the city center of Berlin between 7 and 8 am (cf. Tab. 1).

The total costs to operate the ride-pooling system between 7 and 8 am are composed of the price for the required fuel and the wages of the drivers. From the simulation we get that on average 4688 vehicles, and thus 4688 drivers, are required between 7 and 8am. With an hourly wage of 18€ we get total costs of 84,384€ for the wages between 7 and 8am [24]. According to the simulation, the ride-pooling vehicles drive around 54,000 km during the morning peak. With an average capacity of 32.6kWh/100km and an average price of 30ct per kW/h (assuming that all ride-pooling vehicles are electric vehicles), we get additional costs of 5,281€ for the required energy [25], [26]. Summarized, we get costs of around

90000€ to operate the fleet between 7 and 8 am. If these costs are divided on all users, every trip in the morning would cost around 2.25€ to cover the operational costs, which is comparable to public transport fares. With 251 working days (value for Berlin in 2023 with a 5-day week), we get fare prices of 565€ for every passenger per year.

From the dataset we determine a total length of 138,500 km for the individual car trips between 7 and 8 am in the city center of Berlin. Here, we consider that most of the vehicles are combustion engine vehicles (we only considered diesel oil) with an average consumption of 7 liters per 100km [27]. With an average price of 1.946€ per liter, we get total costs 18,866€ for all trips, resulting in a price of 0.47€ on average for every driver [28]. For 251 working days we thus get yearly costs of 118€ for the trips in the morning. Therefore, using a private vehicle is 80% cheaper than using ride-pooling in this simplified scenario.

This proportion changes if we include the procurement costs. With an average price of 18,800€ per vehicle (only considering pre-owned vehicles), we get total costs of 752 Mil. € for the private vehicles of the 40,000 potential passengers [29].

As price for the vehicles we use 42.690€, the price of a vehicle model often used by ride-pooling operators [30]. For the 4688 vehicles of the fleet we thus get total costs of 200 million euros. Since the vehicles can be used throughout the whole day, only the share of 8.8% has to be covered by the users between 7 and 8 am [19]. This would correspond to costs of 17.6 million euros. Divided on the 40000 users, we get 440€ procurement costs per customer. Thus, the procurement costs customers have to cover are 98% lower for ride-pooling than for the private vehicles.

This comparison is of course highly simplified. For example, the assumption that the introduction of a ride-pooling service will cause customers to give up their private vehicles is rather unrealistic. Furthermore all maintenance costs are completely ignored and for the calculation of the average vehicle price only pre-owned vehicles were taken into consideration. Research into customer price sensitivity has shown that customers tend to underestimate procurement costs and mainly compare fares to fuel costs.. Nevertheless, the comparison provides a basic impression that, considering all costs, ride-pooling is not necessarily more expensive than the large-scale use of private vehicles. It is also obvious that most of the operational costs of a ride-pooling fleet are due to the wages of the drivers. Therefore, autonomous vehicles are an appealing approach to make ride-pooling systems cheaper and, hence, more widely usable.

# 5. Conclusion

We studied in this paper how many trips in the morning peak (between 7 and 8am) in Berlin and in the city center of Berlin could be serviced by a ride-pooling system. We further investigated how this influences the relative trip time of the passengers, the relatively driven distance, the share of empty mileage and the average occupancy of the vehicles. As a database we used real car trips, tracked in 2017, by the company INRIX. To study the ride-pooling service an agent-based ride-pooling simulation is used. We found that in the city center of Berlin it is possible to serve all former car trips with around 4700 vehicles. For the passengers this results in a 1.34 times higher travel time compared to traveling with their own vehicle. The driven distance reduced around 65%, while the average occupancy increased to 2.6. Empty mileage could be ignored, less than 2% of the distance the vehicles drove empty. For the whole area of Berlin we were limited by computational capacities due to the high number of stops and requests. But we were able to show that also for the whole area of Berlin ride-pooling is capable of pooling a high amount of trips and, thus, reducing the driven distance while increasing the average occupancy of the vehicles. With 10000 vehicles in the fleet the driven distance could already be reduced by around 40%, while serving 60% of the former car trips. Simulations of more vehicles would be necessary to find a fleet size, capable of serving all requests in the larger area.

We conclude that the trip density in Berlin would be high enough to warrant efficient ride pooling with acceptable delays as well as competitive fares with human drivers. With autonomous vehicles shared mobility would thus amount to prices at or below the fuel costs of private driving.

As the main problem for the fleet size we recognized the initial positions of the vehicles. We were not able to find an optimal fleet size which is capable of serving all requests while having no unused vehicles. We conclude that rebalancing, in order to get optimal initial positions of the vehicles, is necessary. In the next step we want to compare our results to pure ride-pooling simulation studies in other cities, like Dublin [31].

# Acknowledgments

The authors gratefully acknowledge the European Regional Development Fund (ERDF), the German Federal Ministry of Education and Research and the Land Brandenburg for supporting this project by providing resources on the high performance computer system at the Potsdam Institute for Climate Impact Research.

Alexander Schmaus acknowledges support from the German Federal Environmental Foundation (Deutsche Bundesstiftung Umwelt)

# Literature

[1] IPCC, "Summary for Policymakers," in *Climate Change 2022: Impacts, Adaptation, and Vulnerability. Contribution of Working Group II to the Sixth Assessment Report of the Intergovernmental Panel on Climate Change*, in Climate Change 2022. Cambridge University Press, 2022, p. In Press.

[2] *Umsteuern erforderlich: Klimaschutz im Verkehrssektor*, 1. Auflage, Digitale Originalausgabe. Berlin: Sachverständigenrat für Umweltfragen (SRU), 2017.

[3] EEA, "How air pollution affects our health," May 25, 2023. https://www.eea.europa.eu/en/topics/in-depth/air-pollution/eow-it-affects-our-health (accessed Aug. 27, 2023).

[4] World Economic Forum, "Traffic congestion cost the US economy nearly $87 billion in 2018," Mar. 07, 2023. https://www.weforum.org/agenda/2019/03/traffic-congestion-cost-the-us-economy-nearly-87-billion-in-2018/ (accessed Aug. 27, 2023).

[5] F. Creutzig *et al.*, "SPM5 Demand, Services and Social Aspects of Mitigation".

[6] C. Wilson, L. Kerr, F. Sprei, E. Vrain, and M. Wilson, "Potential Climate Benefits of Digital Consumer Innovations," *Annu. Rev. Environ. Resour.*, vol. 45, no. 1, pp. 113–144, Oct. 2020, doi: 10.1146/annurev-environ-012320-082424.

[7] A. Grubler *et al.*, "A low energy demand scenario for meeting the 1.5 °C target and sustainable development goals without negative emission technologies," *Nat. Energy*, vol. 3, no. 6, pp. 515–527, Jun. 2018, doi: 10.1038/s41560-018-0172-6.

[8] "MOIA." https://help.moia.io/hc/de (accessed Aug. 21, 2023).

[9] A. Schulz, "Rückblick auf den BerlKönig," *Berliner Verkehrsblätter*, pp. 139–144, Jul. 2023.
[10] "Der kommt wie gerufen: Flexible Fahrt mit dem BVG Muva." https://www.bvg.de/de/verbindungen/bvg-muva/flexible-fahrt (accessed Aug. 21, 2023).
[11] R. Tachet *et al.*, "Scaling Law of Urban Ride Sharing," *Sci. Rep.*, vol. 7, no. 1, p. 42868, Mar. 2017, doi: 10.1038/srep42868.
[12] P. Santi, G. Resta, M. Szell, S. Sobolevsky, S. H. Strogatz, and C. Ratti, "Quantifying the benefits of vehicle pooling with shareability networks," *Proc. Natl. Acad. Sci.*, vol. 111, no. 37, pp. 13290–13294, Sep. 2014, doi: 10.1073/pnas.1403657111.
[13] M. H. Chen, A. Jauhri, and J. P. Shen, "Data Driven Analysis of the Potentials of Dynamic Ride Pooling," in *Proceedings of the 10th ACM SIGSPATIAL Workshop on Computational Transportation Science*, Redondo Beach CA USA: ACM, Nov. 2017, pp. 7–12. doi: 10.1145/3151547.3151549.
[14] J. Gödde, L. Ruhrort, V. Allert, and J. Scheiner, "User characteristics and spatial correlates of ride-pooling demand – Evidence from Berlin and Munich," *J. Transp. Geogr.*, vol. 109, p. 103596, May 2023, doi: 10.1016/j.jtrangeo.2023.103596.
[15] F. Jung and D. Manik, "ridepy." Accessed: Aug. 21, 2023. [Online]. Available: https://github.com/PhysicsOfMobility/ridepy
[16] G. Boeing, "OSMnx: New methods for acquiring, constructing, analyzing, and visualizing complex street networks," *Comput. Environ. Urban Syst.*, vol. 65, pp. 126–139, Sep. 2017, doi: 10.1016/j.compenvurbsys.2017.05.004.
[17] "MAPZ," *mapz*. https://www.mapz.com/ (accessed Aug. 21, 2023).
[18] R. Gerike, S. Hubrich, F. Ließke, S. Wittig, and R. Wittwer, "Sonderauswertung zum Forschungsprojekt „Mobilität in Städten–SrV 2018 "," *Städtevergleich Dresd.*, 2020.
[19] R. Gerike, S. Hubrich, F. Ließke, S. Wittig, and R. Wittwer, "Tabellenbericht zum Forschungsprojekt 'Mobilität in Städten-SrV 2018' in Berlin," *TU-Dresd. Dresd.*, 2020.
[20] F. Andre *et al.*, "Digital Auto Report 2021 – Volume 1," 2021.
[21] C. Liebchen, M. Lehnert, C. Mehlert, and M. Schiefelbusch, *Betriebliche Effizienzgrößen für Ridepooling-Systeme*. Springer, 2021.
[22] "Reisezeitindex." https://mobilityinstitute.com/publikationen/reisezeitindex (accessed Aug. 21, 2023).
[23] T. Schlenther, G. Leich, M. Maciejewski, and K. Nagel, "Addressing spatial service provision equity for pooled ride‐hailing services through rebalancing," *IET Intell. Transp. Syst.*, vol. 17, no. 3, pp. 547–556, Mar. 2023, doi: 10.1049/itr2.12279.
[24] BVG, "Gesucht: Busfahrer*innen." https://karriere.bvg.de/berufserfahrene/fahrdienst/busfahrer/busfahrerin-d95 (accessed Aug. 28, 2023).
[25] J. Wieler, "VW T6.1 Elektro von Abt: Warum der Tuner den Elektroantrieb entdeckt hat," May 04, 2023. https://www.adac.de/rund-ums-fahrzeug/autokatalog/marken-modelle/vw/vw-t6-1-elektro/ (accessed Aug. 28, 2023).
[26] Verivox, "Der Strompreis fürs E-Auto." https://www.verivox.de/strom/ratgeber/der-strompreis-fuers-e-auto-1118396/ (accessed Aug. 28, 2023).
[27] Umwelt Bundesamt, "Energieverbrauch und Kraftstoffe," Apr. 27, 2023. https://www.umweltbundesamt.de/daten/verkehr/endenergieverbrauch-energieeffizienz-des-verkehrs#verkehr-braucht-energie (accessed Aug. 28, 2023).
[28] N. Dr. Prack, "Spritpreis-Entwicklung: Benzin- und Dieselpreise seit 1950," Aug. 01, 2023. https://www.adac.de/verkehr/tanken-kraftstoff-antrieb/deutschland/kraftstoffpreisentwicklung/#spritpreise-2011-bis-2022 (accessed Aug. 28, 2023).
[29] J. Wieler and J. Krauß, "Hohe Gebrauchtwagenpreise: Wie man an ein günstiges Auto kommt," Jan. 20, 2023. https://www.adac.de/rund-ums-fahrzeug/auto-kaufen-verkaufen/gebrauchtwagenkauf/gebrauchtwagenmarkt-dat-report/ (accessed Aug. 28, 2023).

[30] VW, "Der Transporter 6.1 Kastenwagen." https://www.volkswagen-nutzfahrzeuge.de/de/modelle/transporter-6-1-kastenwagen.html (accessed Aug. 28, 2023).
[31] K. Jari, "Shared Mobility Simulations for Dublin".